\documentclass[twocolumn,conference]{IEEEtran}
\PassOptionsToPackage{ruled}{algorithm2e}
\usepackage[T1]{fontenc}
\usepackage[latin9]{inputenc}
\usepackage{color}
\usepackage{algorithm2e}
\usepackage{amsmath}
\usepackage{amsthm}
\usepackage{amssymb}
\usepackage{graphicx}

\makeatletter

\providecommand{\tabularnewline}{\\}

\theoremstyle{plain}
\newtheorem{thm}{\protect\theoremname}
\theoremstyle{plain}
\newtheorem{lem}[thm]{\protect\lemmaname}

\usepackage[caption=false,font=footnotesize]{subfig}

\DeclareMathOperator{\maximize}{maximize}

\DeclareMathOperator{\st}{subject~to}
\DeclareMathOperator{\diag}{diag}

\DeclareMathOperator{\tr}{Tr}

\DeclareMathOperator{\vect}{vec}

\newcommand{\herm}{^{{\dagger}}}
\newcommand{\trans}{^{\mathsf{T}}}

\usepackage{cite}
\usepackage{acronym}
\AtBeginDocument{\acrodef{ASP}{antenna separation product}
\acrodef{AWGN}{additive white Gaussian noise}
\acrodef{BEP}{bit error probability}
\acrodef{BER}{bit error rate}
\acrodef{BF-MIMO}[BF\mbox{-}MIMO]{beamforming MIMO}
\acrodef{BF}{beamforming}
\acrodef{bpcu}{bits per channel use}
\acrodef{CP}{cyclic prefix}
\acrodef{CSI}{channel state information}
\acrodef{CSIR}{channel state information at RX}
\acrodef{SSK}{space shift keying}
\acrodef{CSIT}{channel state information at TX}
\acrodef{DCMC}{discrete\mbox{-}input continuous\mbox{-}output memoryless channel}
\acrodef{DFT}{discrete Fourier transform}
\acrodef{DL-TR-GSM}{dual-layered transmit-receive \acl{GSM}}
\acrodef{DLT}{dual-layered transmission}
\acrodef{EGC}{equal gain combining}
\acrodef{EM}{electromagnetic}
\acrodef{FSPL}{free space path loss}
\acrodef{FFT}{fast Fourier transform}
\acrodef{FDE}{frequency domain equalization}
\acrodef{GRSM}{generalized \acl{RSM}}
\acrodef{GSM}{generalized \acl{SM}}
\acrodef{IFFT}{invserse fast Fourier transform}
\acrodef{ICI}{inter-channel interference}
\acrodef{iid}[i.i.d.]{independent and identically distributed}
\acrodef{IQ}{in\mbox{-}phase and quadrature}
\acrodef{ISI}{intersymbol interference}
\acrodef{ISI-free}[ISI\mbox{-}free]{intersymbol interference free}
\acrodef{LIS}{large intelligent surface}
\acrodef{LOS}{line\mbox{-}of\mbox{-}sight}
\acrodef{mmWave}{millimeter-wave}
\acrodef{MIMO}{multiple\mbox{-}input multiple\mbox{-}output}
\acrodef{MISO}{multiple\mbox{-}input single\mbox{-}output}
\acrodef{ML}{maximum likelihood}
\acrodef{MRC}{maximal ratio combining}
\acrodef{MMSE}{minimum mean square error}
\acrodef{MU-TR-GSM}{multiuser transmit-receive  \acl{GSM} }
\acrodef{NCSIT}{no channel state information at TX}
\acrodef{NLOS}{non\mbox{-}\acs{LOS}} 
\acrodef{OFDM}{orthogonal frequency division multiplexing}
\acrodef{PA}{power amplifier}
\acrodef{PAE}{power added efficiency}
\acrodef{PAPR}{peak\mbox{-}to\mbox{-}average power ratio}
\acrodef{PDF}{probability density function}
\acrodef{PEP}{pairwise error probability}
\acrodefplural{PEP}{pairwise error probabilities}
\acrodef{PMP}{probability mass function}
\acrodef{PSM}{precoding-aided spatial modulation}
\acrodef{QSM}{quadrature spatial modulation}
\acrodef{RC}{reorganization computation}
\acrodef{RHS}{right-hand side}
\acrodef{RIS}{reconfigurable intelligent surface}
\acrodef{RSM}{receive spatial modulation}
\acrodef{RX}{receiver}
\acrodef{SEP}{symbol error probability}
\acrodef{SER}{symbol error rate}
\acrodef{SISO}{single-input single-output}
\acrodef{SM}{spatial modulation}
\acrodef{SMX-MIMO}[SMX\mbox{-}MIMO]{spatial multiplexing MIMO}
\acrodef{SMX}{spatial multiplexing}
\acrodef{SNR}{signal-to-noise ratio}
\acrodef{SC}{single carrier}
\acrodef{SVD}{singular value decomposition}
\acrodef{SPST}{single pole single-throw}
\acrodef{TDE}{time domain equalization}
\acrodef{TX}{transmitter}
\acrodef{ULA}{uniform linear array}
\acrodef{URA}{uniform rectangular array}
\acrodef{VGA}{variable gain amplifier}
\acrodef{ZF}{zero-forcing}
\acrodef{ZMCG}{zero-mean complex Gaussian}

}

\usepackage{pgfplots}
\usepackage{grffile}
\pgfplotsset{compat=newest}
\usetikzlibrary{plotmarks}
\usepgfplotslibrary{patchplots}
\usepackage{amsmath}
\usepackage{balance}

\pgfplotsset{compat=newest}
\usetikzlibrary{plotmarks}
\usepgfplotslibrary{patchplots}

\tikzset{every node/.style={font=\small}}
\tikzset{every pin/.style={fill=white,font=\small}}
\tikzset{every pin edge/.style={<-,>=stealth,black,thick}}
\pgfplotsset{grid style={dotted,gray}}
\pgfplotsset{every axis/.style={inner sep=2pt}}
\pgfplotsset{legend style={font=\small}}
\newlength\figurewidth
\newlength\figureheight

\definecolor{mycolor1}{rgb}{1.00000,0.00000,1.00000}%

\makeatother

\providecommand{\lemmaname}{Lemma}
\providecommand{\theoremname}{Theorem}

\begin{document}
\title{Optimization of RIS-aided SISO Systems Based on a Mutually Coupled
Loaded Wire Dipole Model}
\author{\IEEEauthorblockN{Nemanja~Stefan~Perovi\'c\IEEEauthorrefmark{1}, Le-Nam Tran\IEEEauthorrefmark{2},
Marco~Di~Renzo\IEEEauthorrefmark{1}, and Mark~F.~Flanagan\IEEEauthorrefmark{2}}\IEEEauthorblockA{\IEEEauthorrefmark{1}Universit\'e Paris-Saclay, CNRS, CentraleSup\'elec,
Laboratoire des Signaux et Syst\`emes\\
3 rue Joliot Curie, 91192, Gif-sur-Yvette, France\\
Email: marco.di-renzo@universite-paris-saclay.fr}\IEEEauthorblockA{\IEEEauthorrefmark{2}School of Electrical and Electronic Engineering,
University College Dublin\\
Belfield, Dublin 4, D04 V1W8, Ireland\\}}

\maketitle
\begin{abstract}
The \ac{EM} features of \acp{RIS} fundamentally determine their
operating principles and performance. Motivated by these considerations,
we study a \ac{SISO} system in the presence of an \ac{RIS}, which
is characterized by a circuit-based \ac{EM}-consistent model. Specifically,
we model the RIS as a collection of thin wire dipoles controlled by
tunable load impedances, and we propose a gradient-based algorithm
for calculating the optimal impedances of the scattering elements
of the RIS in the presence of mutual coupling. Furthermore, we prove
the convergence of the proposed algorithm and derive its computational
complexity in terms of number of complex multiplications. Numerical
results show that the proposed algorithm provides better performance and converges faster
than a benchmark algorithm.
\end{abstract}

\begin{IEEEkeywords}
Reconfigurable intelligent surface (RIS), thin wire dipole model,
mutual coupling, optimization.\acresetall{}
\end{IEEEkeywords}

\section{Introduction}

\textcolor{black}{\bstctlcite{BSTcontrol}}Reconfigurable intelligent
surfaces (RISs)\acused{RIS} represent a new technology capable of
shaping the radio wave propagation and thereby transforming a generally
unpredictable wireless channel into a controllable and programmable
medium. An \ac{RIS} is typically implemented as a thin metasurface
that consists of low-cost and nearly passive radiating elements that
can be electronically controlled to alter the wavefront of the impinging
waves \cite{di2020smart,pan2022overview}. By optimizing the reflection
coefficients of the RIS elements, it is possible to overcome the adverse
effects of uncontrolled wireless propagation, resulting in a variety
of performance and implementation gains. The RIS technology can be
used to improve the received power \cite{qian2020beamforming}, achievable
rate \cite{perovic2020achievable,wu2019intelligent}, mutual information
\cite{perovic2020optimization,karasik2021adaptive}, interference
management \cite{fu2021reconfigurable} and coverage \cite{yang2020coverage}.
In the aforementioned papers, it is assumed that each RIS element
can reflect any impinging radio wave with unitary power efficiency
and an arbitrary phase shift, and that the \ac{EM} coupling between
adjacent RIS elements is negligible. However, these modeling assumptions
generally overlook the \ac{EM} characteristics and implementation
aspects of RISs \cite{di2022communication}. In order to obtain a
more accurate prediction of the performance of RIS-aided communications,
it is important to utilize RIS models that take into account the EM
properties of the RIS, and to develop efficient optimization algorithms
based on these models.

In \cite{abeywickrama2020intelligent}, the authors introduced a circuit-based
model for the RIS elements, which explicitly accounts for the relationship
between the amplitude and phase of the reflection coefficients. Using
this model, they optimized the achievable rate. A transfer function
for an RIS-aided communication system, where the RIS elements are
modeled as thin wire dipoles driven by tunable impedances, was derived
in \cite{gradoni2021end}. This model accounts for the mutual coupling
among the elements of the RIS, as well as the interactions between
the transmitter, the receiver and the RIS, e.g., if they are in the
near-field of each other. Building upon the model from \cite{gradoni2021end},
an algorithm to maximize the received signal power by optimizing the
impedances of the RIS elements was proposed in \cite{qian2021mutual}.
The algorithm was developed for a \ac{SISO} system, and the authors
utilized an approximation for the transfer function that partially
decouples the interactions between the transmitter, RIS, and receiver
while accounting for the mutual coupling among co-located radiating
elements. The algorithm introduced in \cite{qian2021mutual} was generalized
in \cite{abrardo2021mimo} for optimizing the sum-rate of RIS-assisted
multi-user \ac{MIMO} channels by using the weighted \ac{MMSE} algorithm.
The algorithm introduced in \cite{abrardo2021mimo}
determines the optimal impedances of the RIS elements and the optimal
precoding for each transmitter. Similarly, an \ac{MMSE}-based algorithm
that adjusts the transfer admittance matrix of an RIS in order to
maximize the capacity of a point-to-point RIS-aided \ac{MIMO} system
was introduced in \cite{badheka2021irs}. In \cite{shen2021modeling},
the authors proposed an \ac{EM}-compliant RIS-aided communication
model formulated in terms of scattering parameters. Based on the obtained
circuit-based model, they optimized the scattering matrix of the RIS
with the aim to maximize the received signal power.

The contributions of the present paper are as follows:
\begin{itemize}
\item We consider the circuit-based model introduced in \cite{gradoni2021end}
for an RIS-aided SISO system and aim to maximize the received signal
power for the considered system.
\item We introduce a gradient-based algorithm to find the optimal values
of the tunable impedances of the RIS elements. Unlike the
algorithm in \cite{qian2021mutual}, which is based on an approximation
of the transfer function with unknown impact on the final performance,
the proposed algorithm relies on the exact formulation of the transfer
function.
\item We prove that the proposed algorithm is monotonically convergent and
derive its computational complexity in terms of number of complex
multiplications.
\item We show through numerical simulations that the proposed algorithm
provides similar or better performance than the best-known benchmark
algorithm, while being faster. Also, we demonstrate
that reducing the inter-distance between adjacent RIS elements on
a fixed-size RIS leads to improved performance, as long as the mutual
coupling among the RIS elements is considered in the design.
\end{itemize}
\emph{Notation}: Bold lower and upper case letters represent vectors
and matrices, respectively. $(\cdot)^{*}$ denotes the complex conjugate
operator. $\mathfrak{R}(\mathbf{x})$ and $\mathfrak{I}(\mathbf{x})$
denote the real and imaginary part operators of a vector $\mathbf{x}$,
respectively. $\diag(\mathbf{x})$ denotes a diagonal matrix whose
diagonal entries are given by $\mathbf{x}$. For a matrix $\mathbf{X}$,
$\vect_{d}(\mathbf{X})$ denotes a vector whose elements are the diagonal
elements of $\mathbf{X}$. $\boldsymbol{\mathbf{X}}\trans$ and $\boldsymbol{\mathbf{X}}\herm$
denote the transpose and Hermitian transpose of $\boldsymbol{\mathbf{X}}$,
respectively. $\left\Vert \mathbf{X}\right\Vert $
denotes the Frobenius norm of $\mathbf{X}$ and $|x|$ denotes the
absolute value of $x$. $\nabla_{\mathbf{x}}f(\cdot)$ denotes the
complex gradient of $f(\cdot)$ with respect to $\mathbf{x}$.

\section{System Model and Problem Formulation}

\subsection{System Model}

We consider a communication system comprising of a single-antenna
transmitter and receiver that communicate via an RIS. The RIS consists
of $N_{\mathrm{RIS}}$ reflecting elements arranged in an rectangular
formation and uniformly separated in both dimensions. For ease of
representation, the transmit antenna, the receive antenna, and the
reflecting elements of the RIS are modeled as thin wire dipoles made
of perfectly conducting material. The dipole length, $l$, is assumed
to be much larger than its radius, $a$, i.e., a thin wire approximation
is assumed.

The transmitted signal is generated by a voltage source $V_{\mathrm{G}}$
with an internal impedance $z_{\mathrm{G}}$, which is fed to the
transmit dipole. At the reception, the receive dipole is connected
to the load impedance $z_{\mathrm{L}}$. The voltage at the port of
the receiver is denoted by $V_{\mathrm{L}}$. The EM interaction between
the elements of the RIS is specified by the impedance matrix $\mathbf{Z}_{\mathrm{SS}}\in\mathbb{C}^{N_{\mathrm{RIS}}\times N_{\mathrm{RIS}}}$,
where $\mathbf{Z}_{\mathrm{SS}}(q,p)$ characterizes the mutual coupling
induced by the $p$-th element on the $q$-th element. The elements
(thin wire dipoles) of the RIS are connected to tunable impedances,
denoted by the vector $\mathbf{z}_{\mathrm{RIS}}\in\mathbb{C}^{N_{\mathrm{RIS}}\times1}$.
More precisely, the $s$-th element of $\mathbf{z}_{\mathrm{RIS}}$
represents the tunable impedance connected to the $s$-th RIS element.
 We assume that the tunable impedances $\mathbf{z}_{\mathrm{RIS}}$
have a fixed real part (i.e., the resistance is fixed), which satisfies
the condition $\mathbf{z}_{\mathrm{RIS},\text{Re}}(s)=R_{0}\ge0$
for $s=1,\dots,N_{\text{RIS}}$, where $\mathbf{z}_{\mathrm{RIS},\text{Re}}\triangleq\mathfrak{R}(\mathbf{z}_{\mathrm{RIS}})$,
since the RIS is a nearly-passive device and $R_{0}$ accounts for
the internal losses of the tuning circuits. On the other hand, the
imaginary part of each element of $\mathbf{z}_{\mathrm{RIS}}$, i.e.,
$\mathbf{z}_{\mathrm{RIS},\text{Im}}\triangleq\mathfrak{I}(\mathbf{z}_{\mathrm{RIS}})$,
is an arbitrary real number, whose set of feasible values is $[z_{\min},z_{\max}]$\footnote{This feasible set is chosen to account for all possible imaginary
parts of $\mathbf{z}_{\mathrm{RIS}}$ that can be realized by utilizing
existing tunable components.}, which is to be determined by the proposed optimization algorithm.
For ease of notation, we define the equivalent RIS impedance matrix
as follows:
\begin{equation}
\mathbf{Z}_{\mathrm{SE}}=\mathbf{Z}_{\mathrm{SS}}+\diag(\mathbf{z}_{\mathrm{RIS}})\in\mathbb{C}^{N_{\mathrm{RIS}}\times N_{\mathrm{RIS}}}.
\end{equation}

It was shown in \cite{gradoni2021end} that the end-to-end transfer
function (i.e., the equivalent communication channel) is given by
\begin{equation}
h_{\mathrm{E2E}}=\frac{V_{\mathrm{L}}}{V_{\mathrm{G}}}=\frac{z_{L}\phi_{\mathrm{TR}}}{\tilde{\mathrm{z}}_{\mathrm{T}}\tilde{\mathrm{z}}_{\mathrm{R}}-\phi_{\mathrm{TR}}^{2}}\label{eq:Trans_func-1}
\end{equation}
where $\tilde{\mathrm{z}}_{\mathrm{T}}=z_{\mathrm{G}}+\phi_{\mathrm{TT}}$
and $\tilde{\mathrm{z}}_{\mathrm{R}}=z_{\mathrm{L}}+\phi_{\mathrm{RR}}$,
and the notation
\begin{equation}
\phi_{\mathrm{KL}}=z_{\mathbf{\mathrm{KL}}}-\mathbf{z}_{\mathrm{KS}}\mathbf{Z}_{\mathrm{SE}}^{-1}\mathbf{z}_{\mathrm{SL}}\label{eq:Phi}
\end{equation}
for $\mathrm{K},\mathrm{L}\in\{\mathrm{T},\mathrm{R}\}$ is utilized.
Specifically, $z_{\mathrm{TT}}$ ($z_{\mathrm{RR}}$)
is the self impedance of the transmit antenna (receive antenna), and $z_{\mathrm{TR}}=z_{\mathrm{RT}}$ is
the mutual impedance between the transmit and receive antennas. The
vector $\mathbf{z}_{\mathrm{KS}}\in\mathbb{C}^{1\times N_{\mathrm{RIS}}}$
contains the mutual impedances between the transmit antenna and the
RIS elements if $\text{K}=\text{T}$ or between the receive antenna
and the RIS elements if $\text{K}=\text{R}$; and $\mathbf{z}_{\mathrm{SL}}\in\mathbb{C}^{N_{\mathrm{RIS}}\times1}$
denotes the vector containing the mutual impedances between the RIS
elements and the transmit antenna if $\text{L}=\text{T}$ or between
the RIS elements and the receive antenna if $\text{L}=\text{R}$ \footnote{It is worth noting that the mutual impedances depend on the system
geometry, i.e., the relative position and orientation of the transmitter,
receiver and RIS, which are assumed to be fixed in this work. Also,
finding the optimal position and orientation of the RIS in this context
is an interesting open issue.}.

\subsection{Problem Formulation}

In a SISO system, the most important performance metrics, such as
the channel capacity and the \ac{BER}, are determined by
the signal power at the receive antenna. Therefore, our objective
is to maximize this power by finding the optimal values for the tunable
impedances at the RIS. 

Formally written, the optimization problem is given by
\begin{subequations}
\label{eq:opt_alg}
\begin{align}
\text{\ensuremath{\underset{\mathbf{z}_{\mathrm{RIS}}}{\maximize}}	} & f(\mathbf{z}_{\mathrm{RIS}})=\bigl|h_{\mathrm{E2E}}\bigr|^{2}=\Bigl|\frac{z_{\text{L}}\phi_{\mathrm{TR}}}{\tilde{\mathrm{z}}_{\mathrm{T}}\tilde{\mathrm{z}}_{\mathrm{R}}-\phi_{\mathrm{TR}}^{2}}\Bigr|^{2}\label{eq:Obj_Fun}\\
\st\text{	} & \mathbf{z}_{\mathrm{RIS,}\text{Re}}(s)=R_{0},\thinspace\forall s\\
 & \mathbf{z}_{\mathrm{RIS,}\text{Im}}(s)\in[z_{\min},z_{\max}],\thinspace\forall s.
\end{align}
\end{subequations}

\section{Proposed Optimization Algorithm}

To solve the optimization problem in \eqref{eq:opt_alg}, we propose
an efficient gradient-based method, which iteratively adjusts the
imaginary parts of $\mathbf{z}_{\mathrm{RIS}}$ (i.e., $\mathbf{z}_{\mathrm{RIS,Im}}$).
It is worth noting that this method can be generalized to simultaneously
optimize the real and the imaginary parts of $\mathbf{z}_{\mathrm{RIS}}$ \footnote{The optimization of $\mathbf{z}_{\mathrm{RIS,Re}}$ implies that the
RIS may also consist of active reflecting elements. In this case,
the RIS would no longer be a nearly-passive device and for this reason
we focus solely on optimizing $\mathbf{z}_{\mathrm{RIS,Im}}.$ The
RIS may, however, be still a globally-passive device, as discussed
in \cite{di2022communication}.}. Before presenting the actual optimization algorithm for $\mathbf{z}_{\mathrm{RIS}}$,
we provide the gradient of $f(\mathbf{z}_{\mathrm{RIS}})$ with respect
to $\mathbf{z}_{\mathrm{RIS},\text{Im}}$.
\begin{lem}
\label{lem:Gradient_Zris}The gradient of $f(\mathbf{z}_{\mathrm{RIS}})$
with respect to $\mathbf{z}_{\mathrm{RIS,Im}}$\textup{ }\textup{\emph{is}}
\begin{equation}
\nabla_{\mathbf{z}_{\mathrm{RIS,Im}}}f(\mathbf{z}_{\mathrm{RIS}})=2\mathfrak{I}(h_{\mathrm{E2E}}\vect_{d}(\mathbf{E}^{*})),\label{eq:Gradient_zris}
\end{equation}
where\textcolor{black}{
\begin{align}
\mathbf{E} & =z_{\mathrm{L}}a\mathbf{Z}_{\mathrm{SE}}^{-1}\bigl(\bigl(2a\phi_{\mathrm{TR}}^{2}+1\bigr)\mathbf{z}_{\mathrm{SR}}\mathbf{z}_{\mathrm{TS}}\nonumber \\
 & \quad-a\phi_{\mathrm{TR}}\tilde{\mathrm{z}}_{\mathrm{R}}\mathbf{z}_{\mathrm{ST}}\mathbf{z}_{\mathrm{TS}}-a\phi_{\mathrm{TR}}\tilde{\mathrm{z}}_{\mathrm{T}}\mathbf{z}_{\mathrm{SR}}\mathbf{z}_{\mathrm{RS}}\bigr)\mathbf{Z}_{\mathrm{SE}}^{-1}\label{eq:E_mat}
\end{align}
}and $a=\bigl(\tilde{\mathrm{z}}_{\mathrm{T}}\tilde{\mathrm{z}}_{\mathrm{R}}-\phi_{\mathrm{TR}}^{2}\bigr)^{-1}$.
\end{lem}
\begin{IEEEproof}
See Appendix \ref{sec:Proof_grad_Zris}.
\end{IEEEproof}
Given the gradient in Lemma 1, the proposed optimization algorithm
is based on a line search procedure as detailed in Algorithm 1. Specifically,
the elements of $\mathbf{z}_{\mathrm{RIS}}^{(n+1)}$ at the $(n+1)$-th
iteration of the algorithm are obtained as follows. First, we update
the values of the impedances of the RIS elements based on $\mathbf{z}_{\mathrm{RIS}}^{(n+1)}=R_{0}+jP_{Z}(\mathbf{z}_{\mathrm{\mathrm{RIS},Im}}^{(n)}+\text{\ensuremath{\mu_{n}}}\nabla_{\mathbf{z}_{\mathrm{RIS}\text{,Im}}}f(\mathbf{z}_{\mathrm{RIS}}^{(n)}))$,
where the projection of the impedance $\tilde{\mathbf{z}}_{\mathrm{\mathrm{RIS},Im}}=P_{Z}(\mathbf{z}_{\mathrm{\mathrm{RIS},Im}})$
is given by
\begin{equation}
\tilde{\mathbf{z}}_{\mathrm{\mathrm{RIS},Im}}(s)=\begin{cases}
z_{\max} & \mathbf{z}_{\mathrm{\mathrm{RIS},Im}}(s)>z_{\text{\ensuremath{\max}}}\\
z_{\text{\ensuremath{\min}}} & \mathbf{z}_{\mathrm{\mathrm{RIS},Im}}(s)<z_{\text{\ensuremath{\min}}}\\
\mathbf{z}_{\mathrm{\mathrm{RIS},Im}}(s) & \text{otherwise}
\end{cases}
\end{equation}
for $s\in\{1,2,\dots,N_{\text{RIS}}\}$. Then, we assume that the
gradient step size $\mu_{n}$ is updated based on a line search procedure.
Also, we utilize the following quadratic approximation for the function
$f(\mathbf{z}_{\mathrm{RIS}})$ around $\mathbf{\mathbf{z}}_{\mathrm{RIS}}^{(n)}$:
\begin{multline}
Q_{\text{\ensuremath{\mu_{n}}}}(\mathbf{\mathbf{z}}_{\mathrm{RIS}};\mathbf{\mathbf{z}}_{\mathrm{RIS}}^{(n)})=f(\mathbf{\mathbf{z}}_{\mathrm{RIS}}^{(n)})\\
+\bigl\langle\nabla_{\mathbf{z}_{\mathrm{RIS},\text{Im}}}f\bigl(\mathbf{z}_{\mathrm{RIS}}^{(n)}\bigr),\mathbf{\mathbf{z}}_{\mathrm{RIS,Im}}-\mathbf{\mathbf{\mathbf{z}}}_{\mathrm{RIS,Im}}^{(n)}\bigr\rangle\\
-\frac{1}{2\text{\ensuremath{\mu_{n}}}}\bigl\Vert\mathbf{\mathbf{z}}_{\mathrm{RIS,Im}}-\mathbf{\mathbf{\mathbf{z}}}_{\mathrm{RIS,Im}}^{(n)}\bigr\Vert^{2}\label{eq:Quad_app}
\end{multline}
where $\bigl\langle\mathbf{x},\mathbf{y}\bigr\rangle=\mathbf{x}\trans\mathbf{y}$.
If $Q_{\text{\ensuremath{\mu_{n}}}}(\mathbf{z}_{\mathrm{RIS}}^{(n+1)};\mathbf{\mathbf{\mathbf{z}}}_{\mathrm{RIS}}^{(n)})>f(\mathbf{\mathbf{z}}_{\mathrm{RIS}}^{(n+1)})$
(i.e., a lower bound is not achieved yet), we keep decreasing the
step size $\text{\ensuremath{\mu_{n}}}$ until a quadratic minorant
is achieved. It is worth noting that the proposed procedure terminates
after a finite number of steps, since $f(\mathbf{\mathbf{z}}_{\mathrm{RIS}}^{(n+1)})$
is Lipschitz continuous \cite{beck2009fast}. To speed up the convergence
of Algorithm 1, the step size is readjusted to the initial value after
every 1000 iterations.
\begin{algorithm}[t]
\SetAlgoNoLine
\DontPrintSemicolon
\LinesNumbered  

\KwIn{$\mathbf{z}_{\mathrm{RIS}}^{(0)},\text{\ensuremath{\mu_{\text{init}}}}$}

Set $n\leftarrow0$

Set $\text{\ensuremath{\mu_{0}}\ensuremath{\leftarrow\text{\ensuremath{\mu_{\text{init}}}}}}$

\Repeat{a stopping criterion is met}{

\Repeat{$f(\mathbf{z}_{\mathrm{RIS}}^{(n+1)})\ge Q_{\text{\ensuremath{\mu_{n}}}}(\mathbf{z}_{\mathrm{RIS}}^{(n+1)};\mathbf{\mathbf{z}}_{\mathrm{RIS}}^{(n)})$}{

Set $\mathbf{z}_{\mathrm{RIS}}^{(n+1)}=R_{0}+jP_{Z}(\mathbf{z}_{\mathrm{\mathrm{RIS},Im}}^{(n)}+\text{\ensuremath{\mu_{n}}}\nabla_{\mathbf{z}_{\mathrm{RIS}\text{,Im}}}f(\mathbf{z}_{\mathrm{RIS}}^{(n)}))$

\If{$f(\mathbf{z}_{\mathrm{RIS}}^{(n+1)})<Q_{\text{\ensuremath{\mu_{n}}}}(\mathbf{z}_{\mathrm{RIS}}^{(n+1)};\mathbf{\mathbf{z}}_{\mathrm{RIS}}^{(n)})$
}{$\text{\ensuremath{\mu_{n}}}\leftarrow\kappa\text{\ensuremath{\mu_{n}}}$
\tcc*[f]{$0<\kappa<1$}\;}}

$n\leftarrow n+1$

\If{$\mathrm{mod}(n,1000)=0$ }{$\text{\ensuremath{\mu_{n}}}\leftarrow\text{\ensuremath{\mu_{\text{init}}}}$\;}

}

\caption{Proposed optimization algorithm\label{alg:LS_Zris}}
\end{algorithm}

\section{Convergence and Complexity Analysis}

\subsection{Convergence Analysis}

In this subsection, we analyze the convergence of the proposed algorithm.
Let $L_{\mathbf{z}_{\mathrm{RIS},\text{Im}}}$ be the Lipschitz constant
of $\nabla_{\mathbf{z}_{\mathrm{RIS},\text{Im}}}f(\mathbf{z}_{\mathrm{RIS}})$.
Then
\begin{align}
f(\mathbf{\mathbf{z}}_{\mathrm{RIS}}^{(n+1)}) & \ge f\bigl(\mathbf{\mathbf{z}}_{\mathrm{RIS}}^{(n)}\bigr)+\bigl\langle\nabla_{\mathbf{z}_{\mathrm{RIS},\text{Im}}}f\bigl(\mathbf{\mathbf{\mathbf{z}}}_{\mathrm{RIS}}^{(n)}\bigr),\mathbf{\mathbf{\mathbf{z}}}_{\mathrm{RIS,Im}}^{(n+1)}-\mathbf{\mathbf{z}}_{\mathrm{RIS,Im}}^{(n)}\bigr\rangle\nonumber \\
 & -\frac{L_{\mathbf{z}_{\mathrm{RIS},\text{Im}}}}{2}\bigl\Vert\mathbf{\mathbf{\mathbf{z}}}_{\mathrm{RIS,Im}}^{(n+1)}-\mathbf{\mathbf{z}}_{\mathrm{RIS,Im}}^{(n)}\bigr\Vert^{2}.\label{eq:conv1}
\end{align}

 Since the line search procedure ensures $Q_{\text{\ensuremath{\mu_{n}}}}(\mathbf{z}_{\mathrm{RIS}}^{(n+1)};\mathbf{\mathbf{\mathbf{z}}}_{\mathrm{RIS}}^{(n)})\le f(\mathbf{\mathbf{z}}_{\mathrm{RIS}}^{(n+1)})$,
we have
\begin{align}
\bigl\langle\nabla_{\mathbf{z}_{\mathrm{RIS},\text{Im}}}f\bigl(\mathbf{\mathbf{z}}_{\mathrm{RIS}}^{(n)}\bigr),\mathbf{\mathbf{\mathbf{z}}}_{\mathrm{RIS,Im}}^{(n+1)}-\mathbf{\mathbf{z}}_{\mathrm{RIS,Im}}^{(n)}\bigr\rangle\nonumber \\
-\frac{1}{2\text{\ensuremath{\mu_{n}}}}\bigl\Vert\mathbf{\mathbf{\mathbf{z}}}_{\mathrm{RIS,Im}}^{(n+1)}-\mathbf{\mathbf{z}}_{\mathrm{RIS,Im}}^{(n)}\bigr\Vert^{2} & \ge0\label{eq:conv2}
\end{align}

Combining \eqref{eq:conv1} and \eqref{eq:conv2}, we obtain
\begin{equation}
f(\mathbf{z}_{\mathrm{RIS}}^{(n+1)})\ge f\bigl(\mathbf{z}_{\mathrm{RIS}}^{(n)}\bigr)+\left[\frac{1}{\text{\ensuremath{2\mu_{n}}}}-\frac{L_{\mathbf{z}_{\mathrm{RIS},\text{Im}}}}{2}\right]\bigl\Vert\mathbf{\mathbf{\mathbf{z}}}_{\mathrm{RIS,Im}}^{(n+1)}-\mathbf{\mathbf{z}}_{\mathrm{RIS,Im}}^{(n)}\bigr\Vert^{2}.
\end{equation}

If $\mu_{n}<1/L_{\mathbf{z}_{\mathrm{RIS},\text{Im}}}$, we have $f(\mathbf{\mathbf{z}}_{\mathrm{RIS}}^{(n+1)})>f\bigl(\mathbf{z}_{\mathrm{RIS}}^{(n)}\bigr)$
and hence the objective function is monotonically increasing. 

Also,
the objective function $f(\mathbf{z}_{\mathrm{RIS}})$ is bounded from the above, thanks to the physical nature of the problem being considered. In fact, the transmit power is finite and neither the RIS nor the receiver amplify the signals, since the resistances of the tuning circuits of the RIS and the loads of the receiver are non-negative by design. This property holds true for any feasible set in (4c), including $z_{\min} \to -\infty$ and $z_{\max} \to +\infty$.

Therefore, we conclude that the objective
sequence $\{f(\mathbf{\mathbf{z}}_{\mathrm{RIS}}^{(n)})$\} is provably
convergent.

\subsection{Complexity Analysis}

In this subsection, the computational complexity of the proposed algorithm
is evaluated in terms of number of complex multiplications. The inversion
of the matrix $\mathbf{Z}_{\mathrm{SE}}$ requires $\mathcal{O}(N_{\text{RIS}}^{3})$
multiplications. Additional $\mathcal{O}(N_{\mathrm{RIS}}^{2})$ multiplications
are needed to calculate $\phi_{\mathrm{KL}}$ for $\mathrm{K},\mathrm{L}\in\{\mathrm{T},\mathrm{R}\}$
in \eqref{eq:Phi}. Also, we observe that the complexity of each term
in brackets in \eqref{eq:E_mat} is $\mathcal{O}(N_{\text{RIS}}^{2})$,
and multiplying the central term in brackets with $\mathbf{Z}_{\mathrm{SE}}^{-1}$
on the left-hand and right-hand sides requires $\mathcal{O}(2N_{\text{RIS}}^{3})$
multiplications. Also, the multiplication with $z_{\mathrm{L}}a$
has a complexity of $\mathcal{O}(N_{\text{RIS}}^{2})$. Hence, the
total complexity of calculating the gradient $\nabla_{\mathbf{z}_{\mathrm{RIS}\text{,Im}}}f(\mathbf{z}_{\mathrm{RIS}})$
is approximately $\mathcal{O}(3N_{\text{RIS}}^{3})$. In the inner
loop of Algorithm 1 (Steps 4-9), $\mathcal{O}(N_{\text{RIS}}^{3}+N_{\mathrm{RIS}}^{2})$
multiplications are needed to calculate $f(\mathbf{\bar{z}}_{\mathrm{RIS}}^{(n+1)})$.
The additional complexity of computing $Q_{\boldsymbol{\mu}_{n}}(\mathbf{\bar{z}}_{\mathrm{RIS}}^{(n+1)};\mathbf{\mathbf{\bar{z}}}_{\mathrm{RIS}}^{(n)})$
can be neglected. Hence, the complexity per iteration of the proposed
algorithm (i.e., computing $\mathbf{\bar{z}}_{\mathrm{RIS}}^{(n+1)}$
from $\mathbf{\bar{z}}_{\mathrm{RIS}}^{(n)}$) is $\mathcal{O}(3N_{\text{RIS}}^{3}+L_{P}(N_{\text{RIS}}^{3}+N_{\text{RIS}}^{2})),$
where $L_{P}$ is the average number of inner loops for the line search
procedure. The value of $L_{p}$ can be estimated by performing numerical
simulations based on the proposed method. The (average) total complexity
of the proposed method is given~by 
\begin{equation}
C_{P}=\mathcal{O}(I_{P}(3N_{\text{RIS}}^{3}+L_{P}(N_{\text{RIS}}^{3}+N_{\text{RIS}}^{2}))),
\end{equation}
where $I_{P}$ is the average number of iterations.

\section{Simulation Results}

Wee present numerical results to demonstrate the performance
of the proposed optimization method. As a benchmark, we consider the
optimization algorithm introduced in \cite{qian2021mutual}, which optimizes the signal power at the receive antenna, but uses the
following approximated end-to-end transfer function:
\begin{equation}
h_{\mathrm{E2E}}\approx h_{\mathrm{E2E,APP}}=\mathcal{Y}_{0}(z_{\text{RT}}-\mathbf{z}_{\text{RS}}\mathbf{Z}_{\text{SE}}^{-1}\mathbf{z}_{\text{ST}})=\mathcal{Y}_{0}\phi_{\mathrm{RT}},
\end{equation}
where $\mathcal{Y}_{0}=z_{\text{L}}(z_{\text{L}}+z_{\text{RR}})^{-1}(z_{\text{G}}+z_{\text{TT}})^{-1}$.
In each iteration, the benchmark algorithm updates the RIS impedances
as $\mathbf{\mathbf{z}}_{\mathrm{RIS}}^{(n+1)}=\mathbf{\mathbf{z}}_{\mathrm{RIS}}^{(n)}+j\delta\sin(\boldsymbol{\theta})$,
where $\delta=\mathfrak{R}(\mathbf{Z}_{\text{SS}}(1,1))/M$ and $\boldsymbol{\theta}$
is calculated according to \cite[Eq. (24)]{qian2021mutual}. First,
we present the obtained objective values for the proposed and benchmark
algorithms. Next, we compare the computational complexities and the
execution times of the two algorithms. Finally, we analyze the trade-off
between the inter-distance among the RIS elements and the number of
RIS elements.

The simulation setup considered in this study assumes a frequency
equal to $f=3.5\thinspace\mathrm{GHz}$, whose corresponding wavelength
is $\lambda=8.57\,\mathrm{cm}$. The positions of the transmit antenna,
receive antenna, and the elements of the RIS are specified by Cartesian
coordinates. The RIS has a square shape and is placed on the $yz$-plane
with its center located in $(0,0,0)$. The positions of the transmit
and the receive antennas are $(10\thinspace\mathrm{m},-1\thinspace\mathrm{m},0\thinspace\mathrm{m})$
and $(10\thinspace\mathrm{m},99\thinspace\mathrm{m},0\thinspace\mathrm{m})$,
respectively, resulting in a distance of 100\,m. The number of RIS
elements is $N_{\mathrm{RIS}}=14\times14=196$. The inter-distance
between the RIS elements in both dimensions is $\lambda/4$, and the
RIS size is approximately $30\thinspace\text{cm}\times30\thinspace\text{cm}$. According
to Section II, the transmit and receive antennas, and the reflecting
elements of the RIS are modeled as thin wire dipoles with radius $a=\lambda/500$
and length $l=\lambda/32$, and they are assumed to be placed parallel
to the $z$-axis. Also, we assume $z_{\text{G}}=z_{\text{L}}=(50+50j)\thinspace\text{Ohm}$.
The initial value for the step size in Algorithm 1 is $\mu_{\text{init}}=10^{25}$
and $\kappa=1/2$. The allowed maximum and minimum values for $\mathbf{\mathbf{z}_{\mathrm{RIS},\text{Im}}}$
are $z_{\max}=-z_{\min}=10^{4}\,\text{Ohm}$, respectively. For the
benchmark algorithm, $\mathbf{z}_{\text{RIS}}$ is initialized using
the optimal $\mathbf{z}_{\text{RIS}}$ for the case of no mutual coupling
between the RIS elements \cite[Eq. (10)]{qian2021mutual} and $M=50$.
For the proposed algorithm, the numerical experiments have shown that
the best results are obtained if $\mathbf{z}_{\text{RIS}}$ is initialized
as $\mathbf{z}_{\text{RIS}}^{(0)}=R_{0}-j\mathfrak{I}(\vect_{d}(\mathbf{Z}_{\text{SS}}))$.
Additionally, it is assumed that the direct link is very weak, allowing
us to ignore it, i.e., $z_{\mathrm{TR}}=z_{\mathrm{RT}}=0$. All impedances
are calculated from \cite{gradoni2021end}. The algorithms are developed
using MATLAB R2022a and are run on a computer with a processor whose frequency
is 2.8\,GHz. 
\begin{figure}[t]
\begin{centering}
\includegraphics[width = \columnwidth]{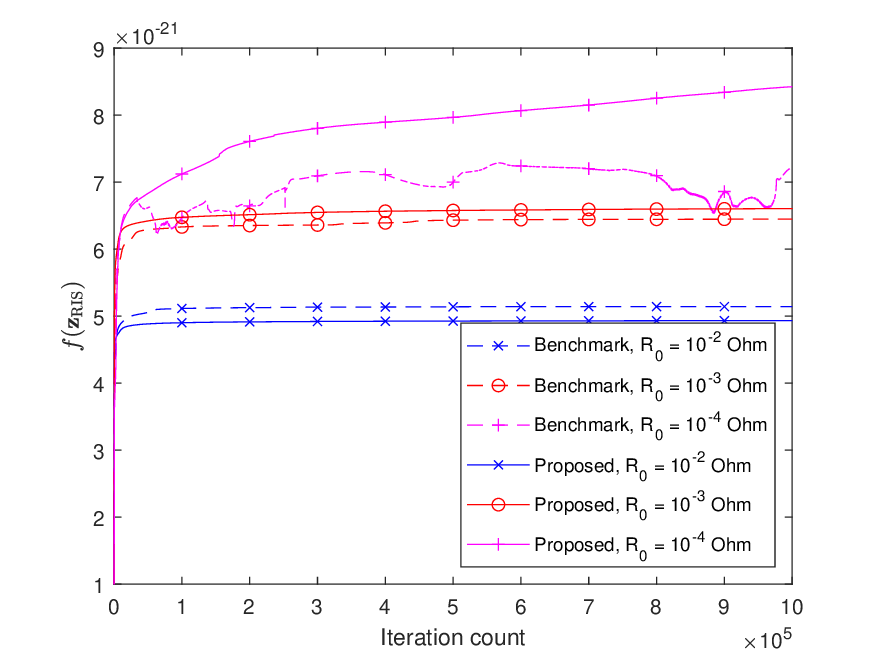}
\par\end{centering}
\centering{}\caption{Objective functions of the proposed and benchmark algorithms versus
the number of iterations. \label{fig:ObjectiveVsResist}}
\end{figure}

Fig. 1 plots the objective functions of the proposed and benchmark
algorithms against the number of iterations. The proposed algorithm
shows a performance advantage compared to the benchmark algorithm
if the real part of $\mathbf{z}_{\mathrm{RIS}}$ is smaller than or
equal to $10^{-3}$ Ohm. Moreover, when the real part of $\mathbf{z}_{\mathrm{RIS}}$
is very small (i.e., $10^{-4}$ Ohm), the benchmark algorithm is not
monotonically convergent. This can be explained by the following argument.
When the real part of $\mathbf{z}_{\mathrm{RIS}}$ is very small,
the algorithm tends to update the imaginary part of $\mathbf{z}_{\mathrm{RIS}}$
so that $\mathbf{Z}_{\mathrm{SE}}$ becomes almost singular. On the
other hand, the impedance increment in an iteration of the benchmark
algorithm in \cite{qian2021mutual} has to satisfy the condition $\delta\ll1/\left\Vert \mathbf{Z}_{\mathrm{SE}}^{-1}\right\Vert $,
which is hard to achieve when $\mathbf{Z}_{\mathrm{SE}}$ tends to
be singular and thus the algorithm is not convergent. The convergence
of the benchmark algorithm can, however, be improved by setting a
very small value of $\delta$, but the corresponding convergence time
significantly increases in this case. The results shown in Fig.~1
are obtained by setting $\delta=0.0039$. In \cite{mursia2023modeling},
the authors tackled this issue by adding the condition $\delta\ll1/\left\Vert \mathbf{Z}_{\mathrm{SE}}^{-1}\right\Vert $
as a constraint of the formulated optimization problem.

\begin{figure}[t]
\includegraphics[width = \columnwidth]{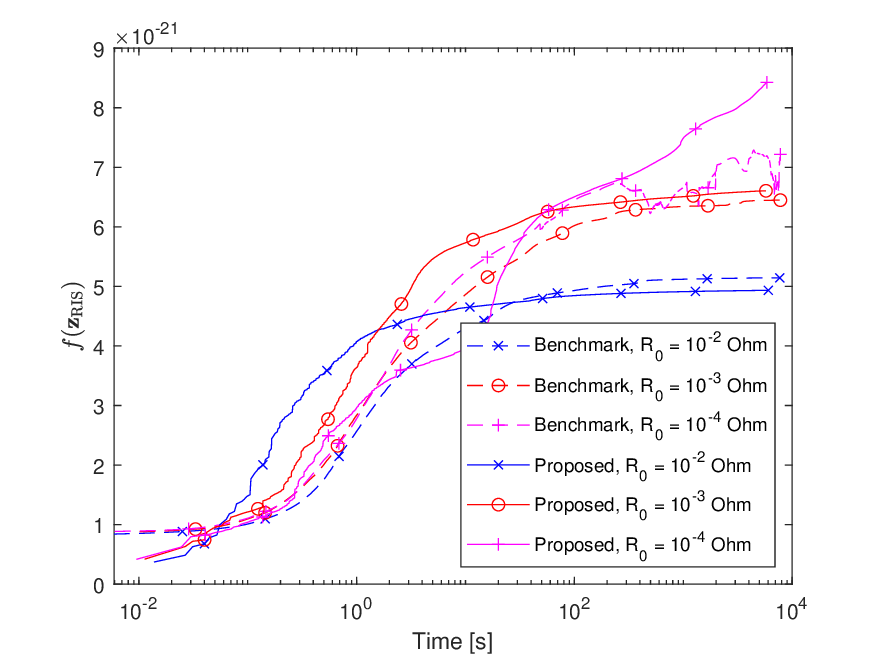}
\centering{}\caption{Objective functions of the proposed and benchmark algorithms versus
the execution time.\label{fig:ObjectiveVsTime}}
\end{figure}

In Fig. \ref{fig:ObjectiveVsTime}, we compare the objective values
of the proposed and benchmark algorithms versus the execution time
using the same simulation setup as in Fig. \ref{fig:ObjectiveVsResist}.
When both algorithms are monotonically convergent, the proposed algorithm
clearly requires less time to converge. As both algorithms optimize
all the RIS tunable impedances in one iteration, the proposed gradient-based
optimization turns out to be more time efficient than the benchmark
algorithm.

The computational complexity required by the proposed and benchmark
algorithms to reach 95\% of the objective values at the $10^{6}$-th
iteration is presented\footnote{We do not present the computational complexity for $R_{0}=10^{-4}$
Ohm, since the benchmark algorithm is not monotonically convergent
for the considered value of $\delta$, and it is difficult to determine
the computation time.} in Table \ref{tab:Comparison-of-the}. The computational complexity
of the benchmark algorithm is $C_{B}=\mathcal{O}(I_{B}(N_{\text{RIS}}^{3}+N_{\text{RIS}}^{2}))$,
where $I_{B}$ is the number of iterations to reach convergence. The
proposed algorithm needs fewer iterations to achieve the target value
of the objective function compared to the benchmark algorithm. However,
due to multiple inner loops in the line search procedure, the proposed
algorithm has a higher computational complexity compared to the benchmark
algorithm. In general, we conclude that the proposed algorithm has
a higher computational complexity, but the execution time is shorter.
\begin{table}[t]
\caption{Comparison of the computational complexity required by the proposed
algorithm ($C_{P}$) and benchmark algorithm ($C_{B}$) to achieve
95\% of the objective values at the $10^{6}$-th iteration.\label{tab:Comparison-of-the}}

\centering{}\resizebox{\columnwidth}{!}{%
\begin{tabular}{|c|c|c|c|c|c|}
\hline 
$R_{0}$ {[}Ohm{]} & $L_{P}$ & $I_{P}$ & $C_{P}$ & $I_{B}$ & $C_{B}$\tabularnewline
\hline 
\hline 
$10^{-2}$ & 6 & 3208 & $1.94\times10^{11}$ & 9683 & $7.32\times10^{10}$\tabularnewline
\hline 
$10^{-3}$ & 6 & 10935 & $6.61\times10^{11}$ & 22138 & $1.68\times10^{11}$\tabularnewline
\hline 
\end{tabular}}
\end{table}

To analyze the execution time in more detail, Table \ref{tab:Time-Table}
presents the time required by the proposed algorithm ($T_{P}$) and
benchmark algorithm ($T_{B}$) to achieve 95\% of the objective values
at the $10^{6}$-th iteration. The results indicate that the proposed
algorithm generally requires considerably less time to achieve the
target objective function value. 
\begin{table}[t]
\centering{}\caption{Time required by the proposed algorithm ($T_{P}$) and benchmark algorithm
($T_{B}$) to achieve 95\% of the objective values at the $10^{6}$-th
iteration.\label{tab:Time-Table}}
\begin{tabular}{|c|c|c|}
\hline 
$R_{0}$ {[}Ohm{]} & $T_{P}$ {[}ms{]} & $T_{B}$ {[}ms{]}\tabularnewline
\hline 
\hline 
$10^{-2}$ & 16 & 67\tabularnewline
\hline 
$10^{-3}$ & 62 & 173\tabularnewline
\hline 
\end{tabular}
\end{table}
\begin{figure}[t]
\raggedright{}\includegraphics[width = \columnwidth]{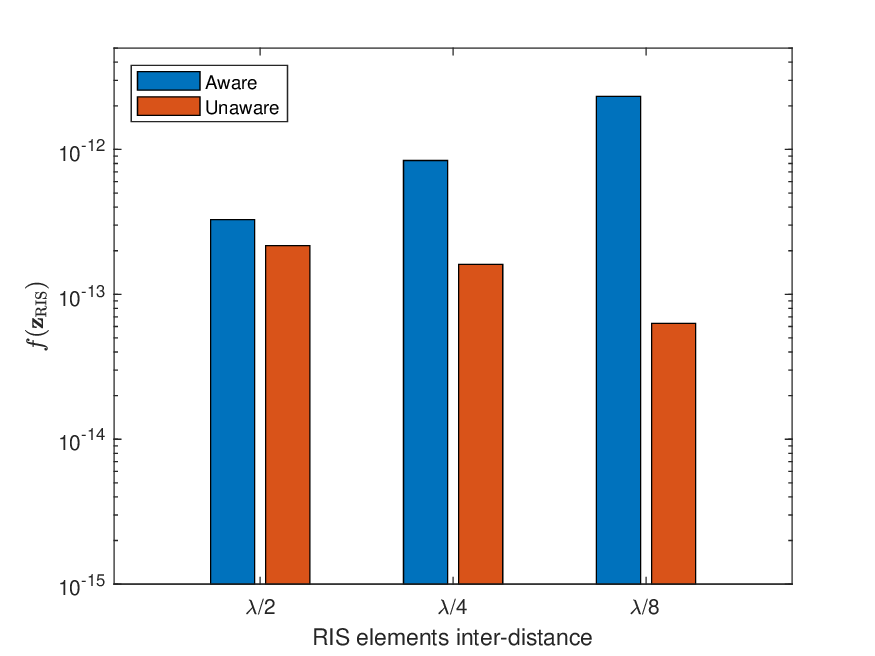}\caption{Objective function of the proposed algorithm versus the inter-distance
between adjacent RIS elements for an RIS of fixed size ($R_{0}=10^{-3}$
Ohm). \label{fig:Obj_RIS-elem_Separ}}
\end{figure}

In Fig. \ref{fig:Obj_RIS-elem_Separ}, we show the objective values
of the proposed algorithm as a function of the inter-distance between
adjacent RIS elements while keeping the size of the RIS fixed. As
a result, the number of RIS elements varies with the inter-distance
between the RIS elements. For an RIS whose size is $15\times15\:\text{cm}$
and the inter-distance is $\lambda/2$, $\lambda/4$ and $\lambda/8$,
the number of RIS elements is 16, 49, and 196, respectively. The real
part of each element of $\mathbf{z}_{\mathrm{RIS}}$ is $R_{0}=10^{-3}$
Ohm. Also, we assume that the lengths of the transmit and receive
antenna dipoles are $\lambda/2$. Simulation results are generated
for the mutual coupling aware and the mutual coupling unaware cases\footnote{In the mutual coupling aware case, the proposed algorithm takes the
mutual coupling between the RIS elements into account at the optimization
stage (i.e., by design). On the other hand, the mutual coupling is
not considered (i.e., ignored) by the proposed algorithm in the mutual
coupling unaware case.}. Similar to \cite[Fig. 3]{qian2021mutual} and \cite{abrardo2021mimo},
the objective function monotonically increases with the number of
RIS elements in the mutual coupling aware case, while it monotonically
decreases in the mutual coupling unaware case. This behavior implies
that taking the mutual coupling into account at the design stage can
enhance the system performance\footnote{These results are obtained assuming that the number of
RIS elements increases as the inter-distance decreases (the
impact of the mutual coupling become more noticeable). This is different
from the results reported in \cite[Fig. 2]{qian2021mutual}, where
the inter-distance decreases but the number of RIS elements is
fixed. In \cite[Fig. 2]{qian2021mutual}, we see a slight decrease
of the objective function as the inter-distance decreases, even in
the mutual coupling aware case.}. Also, the larger values of the objective function with respect to
Fig.~1 are attributed to the larger size of the transmit and receive
antenna dipoles.

\begin{figure}[t]
\raggedright{}\includegraphics[width = \columnwidth]{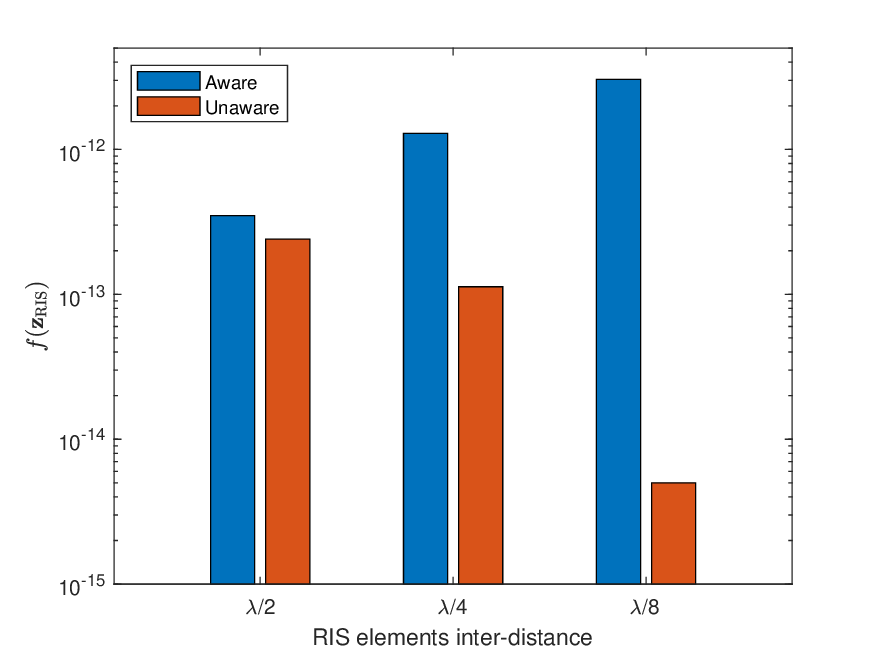}\caption{Objective function of the proposed algorithm versus the inter-distance
between adjacent RIS elements for an RIS of fixed size and for a variable
length of the RIS elements ($R_{0}=10^{-3}$ Ohm). \label{fig:Obj_Val}}
\end{figure}
Fig. \ref{fig:Obj_Val} reports results from a setup analogous to
that of Fig. \ref{fig:Obj_RIS-elem_Separ}, with the exception that
the length of the RIS (dipole) elements is equal to the inter-distance
between them. In the mutual coupling aware case, the objective function
is slightly larger than Fig. \ref{fig:Obj_RIS-elem_Separ}. When the
mutual coupling is disregarded, on the other hand, the objective function
has a significantly lower value. This is attributed to the more profound
impact of the mutual coupling, due to the increased size of the RIS
elements, if it is not taken into account by~design.

\section{Conclusion}

In this paper, we considered a mutually coupled loaded wire dipole
model for RIS-aided communications and proposed a gradient-based algorithm
for optimizing the impedances of the RIS elements. We proved that
the proposed algorithm is monotonically convergent and derived its
computational complexity. Our numerical results showed that the proposed
algorithm achieves better performance compared to a recent benchmark
algorithm and, at the same time, requires significantly less time
to converge. Also, we demonstrated that reducing the inter-distance
between adjacent RIS elements, by keeping the size of the RIS fixed,
positively impacts the system performance, provided that the mutual
coupling is incorporated into the design during the optimization stage.

\appendices{}

\section{\label{sec:Proof_grad_Zris}Proof of Lemma \ref{lem:Gradient_Zris}}

Differentiating the objective function, we have\textcolor{black}{
\begin{equation}
\text{d}f(\mathbf{z}_{\mathrm{RIS}})=\text{d}\bigl|h_{\mathrm{E2E}}\bigr|^{2}=h_{\mathrm{E2E}}^{*}\text{d}h_{\mathrm{E2E}}+h_{\mathrm{E2E}}\text{d}h_{\mathrm{E2E}}^{*}.
\end{equation}
After differentiating the transfer function $h_{\mathrm{E2E}}$ and
performing some simple algebraic manipulations, we obtain
\begin{align}
\text{d}h_{\mathrm{E2E}} & =z_{L}a\bigl(\bigl(2a\phi_{\mathrm{TR}}^{2}+1\bigr)\text{d}\phi_{\mathrm{TR}}\nonumber \\
 & -a\phi_{\mathrm{TR}}\tilde{\mathrm{z}}_{\mathrm{R}}\text{d}\tilde{\mathrm{z}}_{\mathrm{T}}-a\phi_{\mathrm{TR}}\tilde{\mathrm{z}}_{\mathrm{T}}\text{d}\tilde{\mathrm{z}}_{\mathrm{R}}\bigr),
\end{align}
where we use the following identity $\text{d}\phi_{\mathrm{KL}}=\mathbf{z}_{\mathrm{KS}}\mathbf{Z}_{\mathrm{SE}}^{-1}\text{d}\bigl(\diag(\mathbf{z}_{\mathrm{RIS}})\bigr)\mathbf{Z}_{\mathrm{SE}}^{-1}\mathbf{z}_{\mathrm{SL}}$
}for $\mathrm{K,L\in\{T,R\}}$\textcolor{black}{.}

\textcolor{black}{Considering the definition of $\mathbf{E}$ in \eqref{eq:E_mat},
we can write
\begin{align}
\text{d}h_{\mathrm{E2E}} & =\tr\bigl(\mathbf{E}\text{d}\mathbf{z}_{\mathrm{RIS}}\bigr)=\bigl(\vect_{d}(\mathbf{E})\bigr)\trans\text{d}\mathbf{z}_{\mathrm{RIS}},
\end{align}
and thus
\begin{multline*}
\!\!\!\!\!\!\text{d}f(\mathbf{z}_{\mathrm{RIS}})=h_{\mathrm{E2E}}^{*}\bigl(\vect_{d}(\mathbf{E})\bigr)\trans\text{d}\mathbf{z}_{\mathrm{RIS}}+h_{\mathrm{E2E}}\bigl(\vect_{d}(\mathbf{E}^{\ast})\bigr)\trans\text{d}\mathbf{z}_{\mathrm{RIS}}^{*}.
\end{multline*}
Substituting $\text{d}\mathbf{z}_{\mathrm{RIS}}=j\text{d}\mathbf{z}_{\mathrm{RIS},\text{Im}}$
into the previous expression, we obtain
\begin{align}
\text{d}f(\mathbf{z}_{\mathrm{RIS}}) & =2\mathfrak{I}(h_{\mathrm{E2E}}\bigl(\vect_{d}(\mathbf{E}^{\ast})\bigr)\trans)\text{d}\mathbf{z}_{\mathrm{RIS},\text{Im}}.
\end{align}
Finally, we have}
\begin{equation}
\nabla_{\mathbf{z}_{\mathrm{RIS},\text{Im}}}f(\mathbf{z}_{\mathrm{RIS}})=2\mathfrak{I}(h_{\mathrm{E2E}}\vect_{d}(\mathbf{E}^{\ast})),
\end{equation}
\textcolor{black}{which completes the proof.}

\section*{Acknowlegment}

This work was supported in part by the European Commission through
the H2020 SURFER project (grant 101030536), the H2020 ARIADNE
project (grant 871464), the H2020 RISE-6G project (grant 101017011), and by the Agence Nationale de la Recherche (ANR PEPR-5G and Future Networks, grant NF-PERSEUS, 22-PEFT-004).

\bibliographystyle{IEEEtran}
\bibliography{IEEEabrv,references,IEEEexample}

\end{document}